\RequirePackage[2020-02-02]{latexrelease}
\documentclass[preprint,keywords,superscriptaddress,amssymb,nofootinbib,aps]{revtex4}
\usepackage{array,multirow}
\usepackage{mathrsfs}
\usepackage{amsmath}
\usepackage{amsthm}
\usepackage{amssymb, physics}
\usepackage{mathrsfs}
\usepackage{amsmath}
\usepackage{amsthm}
\usepackage{amssymb}

\def\bee{\begin{equation}}
\def\eee{\end{equation}}
\def\baa{\begin{eqnarray}}
\def\eaa{\end{eqnarray}}

\begin{document}

\title{\textbf{Restoring the gauge invariance in non-Abelian second-class theories}} 
 
\author{Paulo R. F. Alves}\email{paulo.alves@ice.ufjf.br}
\affiliation{Departamento de F\'{i}sica, Universidade Federal de Juiz de Fora, 36036-330, Juiz de Fora, MG, Brazil}	
\author{Cleber N. Costa}\email{cleber.costa@ice.ufjf.br}
\affiliation{Departamento de F\'{i}sica, Universidade Federal de Juiz de Fora, 36036-330, Juiz de Fora, MG, Brazil}
\author{Diego Fiorentini}\email{diego_fiorentini@if.uff.br}
\affiliation{UPN - Universidad Privada del Norte, Departamento de Ciencias, 13007, Trujillo, La Libertad, Perú.}
\affiliation{Instituto de F\'isica, Universidade Federal Fluminense, Av. Litor\^anea, s/n, 24210-346, Niter\'oi, RJ, Brasil}                                     
\author{Victor J. Vasquez~Otoya}\email{victor.vasquez@ifsudestemg.edu.br}  
\affiliation{Instituto Federal de Educac\~ao, Ci\^encia e Tecnologia do Sudeste de Minas Gerais,
 Rua Bernardo Mascarenhas 1283, 36080-001, Juiz de Fora, MG, Brasil}
\author{Everton M. C. Abreu}\email{evertonabreu@ufrrj.br}
\affiliation{Departamento de F\'{i}sica, Universidade Federal Rural do Rio de Janeiro, 23890-971, Serop\'edica, RJ, Brazil}
\affiliation{Departamento de F\'{i}sica, Universidade Federal de Juiz de Fora, 36036-330, Juiz de Fora, MG, Brazil}
\affiliation{Programa de P\'os-Gradua\c{c}\~ao Interdisciplinar em F\'isica Aplicada, Instituto de F\'{i}sica, Universidade Federal do Rio de Janeiro, 21941-972, Rio de Janeiro, RJ, Brazil}
\author{Jorge Ananias Neto}\email{jorge@fisica.ufjf.br}
\affiliation{Departamento de F\'{i}sica, Universidade Federal de Juiz de Fora, 36036-330, Juiz de Fora, MG, Brazil}


\begin{abstract}
In this paper, we propose a generalization of an improved gauge unfixing formalism in order to generate gauge symmetries in the non-Abelian valued systems. This generalization displays a proper and formal reformulation of second-class systems within the phase space itself. We then present our formalism in a manifestly gauge invariant resolution of the $SU(N)$ massive Yang-Mills and $SU(2)$ Skyrme models where gauge invariant variables are derived allowing then the achievement of Dirac brackets, gauge invariant Hamiltonians and first-class Lagrangians.
\end{abstract}

\maketitle

\section{Introduction}

Constrained dynamical systems have among other important features the existence of first and second-class constraints \cite{dirac1933, Dirac:1950pj, Dirac:1951zz, Dirac:1958sq, Henneaux:1992ig}. A constraint is said
first-class if its Poisson brackets with all constraints vanish weakly. A constraint that is not first-class is called second-class. First-class constraints imply the existence of a gauge invariance and it is well known that
first-class constraints can be used to generate gauge transformations. gauge symmetries are important in the context of the modern fundamental interaction theories and gravitation. Their rich underlying structures allow us to establish currents and charges conserved (conservation laws) and then defining physical states by removing unphysical degrees of freedom and preserving the unitarity of the S-matrix elements, as well as to establish the Ward identities in order to prove the renormalizability and to remove anomalies. In fact, the principle of gauge symmetry along with generalized quantum field theories based on the standard mathematical techniques for perturbation theories (see, for instance, \cite{Ojima:1978hy,Kugo:1979gm,Bertlmann:295092,Cornwall:1981zr}) can be used as an alternative approach to construct physical theories. In addition it provides a natural geometric-topological set-up for describe a wide range of quantization aspects on pure algebraic grounds by using the homotopy (or (co-)homology) operators \cite{Henneaux:1992ig, Bertlmann:295092}.

Consequently, any fundamental theory can be better structured and understood if formulated as a genuine gauge theory since the advantages are many. A prototypical example of 
second-class constrained dynamical system are the massive Yang-Mills theories. At the original construction, the formation of a dynamical gluon mass $m^2A_\mu^a A_\mu^a$ is forbidden due to the fact that it destroys the gauge invariance. However, there are powerful reasons to consider a mass term since it might play an important role in the nonperturbative regime. This possibility was pointed out by the considerable amount of results obtained through theoretical and phenomenological studies (as in Gribov-Zwanziger, Schwinger-Dyson or QCD sum rules models) as well as from lattice simulations (\textit{vide}, for instance, \cite{Cornwall:1981zr,Greensite:1985tom,Lavelle:1988eg,Gubarev:2000nz,Kondo:2001nq,Dudal:2003vv,Boucaud:2001st,Pene:2010wtc,Aguilar:2015bud} and references therein). Then, this is a clear example where it is desirable to reformulate the mass term to restore the original gauge symmetry. Nevertheless, at low energy QCD standard perturbative techniques cannot be used due to the coupling constant is strong. Effective degrees of freedom in QCD are given by hadrons and another exotic composite states, so the study of hadronic properties requires most of the times of effective models\footnote{Numerical calculations in lattice QCD suffers tecnical problems to implement hadron proprieties too due to the chiral symmetry breaking.}. A particular form to describe baryons and its interactions with a medium is given by the Skyrme model \cite{Skyrme:1961vq}, a nonlinear mesonic model with non-trivial soliton field (who acts as baryon). The Skyrme term is a chiral perturbative correction to the non-linear $\sigma$-model which gives rise to the second-class constraints. It has a great versatility to model diverse problems of nuclear matter among which we can mention spin-isospin correlations (axial coupling), charge radii, magnetic moments, many-bodies nuclear interaction, hadronic crystal lattice and medium modified meson proprieties \cite{Adkins:1983ya,Klebanov:1985qi,Kugler:1989uc,Lee:2003aq}. For a complete review see \cite{Brown:2009eh}. 

The gauge freedom can result in a simplified algebras by choosing the adequate fixing term in each case\footnote{Several algebraic consequences of the first-class conversion were explored in \cite{Banerjee:1993ae,Oliveira:1998ek}.}. The general idea is to rewrite a second-class system as a first-class one, and thus the complete dynamical system exhibits a gauge symmetry. Several methods in the Hamiltonian formalism are available to implement this process among which we can mention the Batalin-Fradkin-Tyutin \cite{Batalin:1986aq, Batalin:1991jm, Banerjee:1993ae, Oliveira:1998ek} and the gauge unfixing (GU) formalism  \cite{Mitra:1990mp,Anishetty:1992yk, Vytheeswaran:1994np}. The last formalism performs this conversion to equivalent gauge invariant theories without the extension of the original phase space variables. The original motivation of the GU formalism and its improved versions \cite{Neto:2006gt,Neto:2009rm} is to use one of the second-class constraint\footnote{in case the system has two second-class constraints.} as a symmetry generator. The other will be discarded. Then, the invariant quantities obtained will be written as a powers series of the discarded constraint. The coefficients of this series will be given by successive transformations of the original variabe under the new symmetry generator \cite{Vytheeswaran:1994np, Vytheeswaran:1997jr}. In quantum field theory, this formalism has been successfully  used in Abelian models such as Chern-Simons, Proca and Carroll-Field-Jackiw \cite{Vytheeswaran:1994np, Vytheeswaran:1997jr, Alves:2020pty}. Then, the aim of the present work
is to face the issue of the gauge invariance restitution of non-Abelian second-class theories. As far as we know, the only attempt to approach a non-abelian model using the GU formalism was performed in \cite{doi:10.1142/S021773231450028X} for pure non-Abelian Chern-Simons theory (without Maxwelll term). Here we provide a general analysis of the GU variables as a power series of the discarded constraint. In all the above mentioned articles related to the use of GU formalism, the correction terms in order $n\geq 2$ are null in the power series including the non-Abelian Chern-Simons theory. This will not be the case for Yang-Mills theories as we will show later. In fact, the issue of the gauge invariance for non-Abelian vector fields has been addressed in \cite{Lavelle:1995ty} where the dressing functions written in trems of infinite sum of non-local terms are used to define invariant quantities. Here we can mention that, for the non-Abelian Stückelberg mechanism with mass term, the invariance requeriment implies a non-polynomial action \cite{Dragon:1996tk,Capri:2016ovw,Capri:2016jgn}. Therefore, it is natural to be expected that, unlike the Abelian cases, Yang-Mills models, whose gauge transformations depend on the covariant derivative, will generate invariant variables without truncations in the power series.

The paper is organized as follows: In Sec. II we give an outline of the improved GU formalism and its non-Abelian extension. In Sec. III we apply our formalism  to the $SU(N)$ massive Yang-Mills model. Sec IV is devoted to the study of the $SU(2)$ Skyrme model. Sec. V contains our conclusions.

\section{The Improved Gauge Unfixing Formalism for Non-Abelian Cases}
\label{GeneralGU}

Let us consider a constrained dynamical system possessing two second class constraints, $T_1^a$ and $T_2^a$, with canonical variables valued in a Lie algebra with $a=1,2,\ldots,N$, where $N$ is the dimension of the representation of these fields with respect to the non-Abelian group. By the definition, the Poisson bracket of these constraint is non non-trivial and it is given by
\begin{equation}
    \left\{T_1^a,T_2^b\right\}=\delta^{ab} K,
\end{equation}
 where $K\neq0$ is non-vanishing constant on the surface defined by the constraints  $T_1^a$ and $T_2^a$. Following the gauge unfixing formalism \cite{Anishetty:1992yk,Vytheeswaran:1994np}, one of the constraints  can be used as generator of infinitesimal transformations and the other one is disregarded (no longer considered a constraint after the reformulation). Without any loss of generality, we can also set the generator of infinitesimal transformations as\footnote{The symbol ``$\approx$" means weak equality, namely they holds only on the hyper-surface defined by the intersection of the constraints on the full phase space.}
\begin{equation}
    \chi^a=K^{-1}T^a_1\approx 0,
\end{equation}
such that, on the surface defined only by $\chi^a$,
\begin{align}
    \left\{\chi^a,T_2^b\right\}&\approx\left\{K^{-1}T_1^a,T_2^b\right\}\nonumber\\
    &\approx K^{-1}\left\{T_1^a,T_2^b\right\}+\left\{K^{-1},T_2^b\right\}T_1^a\nonumber\\
    &\approx \delta^{ab}+\left\{K^{-1},T_2^b\right\}K\chi^a,
\end{align}
that is, $\chi^a$ and $T_2^a$ form an approximate canonically conjugate pair on $\chi^a\approx0$. The constraint $T_2^a$ will be disregarded, that is, it will no longer be considered a constraint. Consider now a second class function $F^a(q^a,p^a)$ \footnote{The ``color" index ``a,b,c..." represents the degrees of freedom and take values of fields in the Lie Algebra }. In the improved gauge unfixing formalism \cite{Neto:2006gt,Neto:2009rm}, the invariant first class function $\tilde{F}^a$ is constructed by as a power series of the discarded constraint $T_2^a$:
\begin{align}
    \tilde{F^a}&=F^a+c_1^{ab}T_2^b+c_2^{abc}T_2^bT_2^c+\ldots\nonumber\\
    &=F^a+\sum_{k=1} c_k^{aw_1\ldots w_k}\prod_{i=1}^kT_2^{w_i},
    \label{F-tilde_1}
\end{align}
which has the following boundary condition 
\begin{equation}
\tilde{F}^a|_{T_2^a\approx 0}=F^a.
\end{equation}
The coefficients $c_k^{aw_1\ldots w_k}$ in \eqref{F-tilde_1} are determined order by order in powers of $T_2^a$ if we employ the invariance condition:
\begin{align}
    \delta^m \tilde{F}^a&=\varepsilon\left\{\tilde{F}^a,\chi^m\right\}=0\\
    &=\delta^mF^a+\sum_{k=1}\left(\delta^mc_k^{aw_1\ldots w_k}\prod_{i=1}^kT_2^{w_i}+c_k^{aw_1\ldots w_k}\prod_{i=1}^k\delta^m T_2^{w_i}\right)=0,
    \label{F-var}
\end{align}
where
\begin{align}
    \delta^m F^a&=\varepsilon\left\{F^a,\chi^m\right\}\\
    \delta^m c_k^{aw_1\ldots w_k}&=\varepsilon\left\{c_k^{aw_1\ldots w_k},\chi^m\right\}\\
    \delta^m T_2^a&=\varepsilon\left\{T_2^a,\chi^m\right\}=-\delta^{am}\varepsilon.
    \end{align}
In the above equations,  $\delta^a$ is the variational operator valued in the Lie algebra for the transformation induced by the generator $\chi^a$ and $\varepsilon$ is an arbitrary infinitesimal parameter. We will assume that there is no an \textit{a priori} symmetry in the color indices of the coefficients $c_k$. Then, from \eqref{F-var}, we can derive an equation for zeroth order terms in $T_2^a$:
\begin{equation}
\delta^m F^a+c_1^{ab}\delta^m T_2^b=0
\Rightarrow \quad c_1^{am}=\frac{\delta^m F^a}{\varepsilon}
\label{con-1}
\end{equation}
For linear terms in $T_2^a$, we get
\begin{equation}
    \delta^mc_1^{ab}T_2^b+c_2^{abc}\delta^mT_2^bT_2^c+c_2^{abc}T_2^b\delta^mT_2^c=0
     \Rightarrow \quad c_2^{amb}=\frac{\delta^m\delta^bF^a}{2\varepsilon^2}
     \label{con-2}
\end{equation}
Then, from equations \eqref{con-1} and \eqref{con-2}, we can observe that, for $k\geq3$, the general relation is
\begin{equation}
    c_{k}^{aw_1\ldots w_k}=\frac{1}{k!\varepsilon^k}\prod_{i=1}^k\delta^{w_i}F^a,
    \label{con-n}
\end{equation}
This expression is the non-Abelian analogue of the expression found in \cite{Neto:2009rm,doi:10.1142/S021773231450028X,Alves:2020pty}. Therefore, the expression for the invariant first class function $\tilde F^a$, eq. \eqref{F-tilde_1}, becomes
\begin{align}
\tilde{F^a}&=F^a+T_2^b\frac{\delta^bF^a}{\varepsilon}+T_2^bT_2^c\frac{\delta^b\delta^cF^a}{2\varepsilon}+\ldots\nonumber\\
&=\left(1+\sum_{k=1}\frac{1}{k!\varepsilon^k}\prod_{i=1}^kT_2^{w_i}\delta^{w_i}\right)F^a\nonumber\\
&=:e^{T_2^b\frac{\delta^b}{\varepsilon}}:F^a,
\label{F-tilde_2}
\end{align}
where we adopted an ordering prescription $``::"$ that $T_2^a$'s always come before the variation operator
\begin{equation}
 \frac{1}{\varepsilon}\delta^af=\left\{f,\chi^a\right\},   
\end{equation}
for any functional $f$ on the phase space. The last line in \eqref{F-tilde_2} shows us that, as pointed out by \cite{Vytheeswaran:1994np}, any gauge invariant quantity can be generated by applying the projection operator $\mathcal{P}=:e^{T_2^b\frac{\delta^b}{\varepsilon}}:$ on function in the phase space. 
In the next section, we will consider some examples to show the use of the improved gauge unfixing formalism for some non-Abelian theories.

\section{Massive $SU(N)$ Yang-Mills model}

To start the illustrations, we begin with massive $SU(N)$ Yang-Mills model in $d=4$ Euclidean dimensions\footnote{We are working in Euclidean space to avoid issues of the validity of the Wick rotation at the non-perturbative regime.}, which is represented by the following action:
\begin{equation}
S = 
\int d^4x \left(\frac{1}{4}\,F^{a}_{\mu\nu}F^a_{\mu\nu}
+\frac{m^{2}}{2}\,A^a_{\mu}A^a_{\mu}\right)  \;, \label{act1}
\end{equation} 
where $F_{\mu\nu}^a=\partial_{\mu}A_{\nu}^a-\partial_{\nu}A_{\mu}^a+f^{abc}A_{\mu}^bA_{\nu}^c$ is the field strength. In other words, this is the non-Abelian version of the Proca Lagrangian. Because of the presence mass term, this Lagrangian is no longer invariant under gauge transformations. The canonical momenta are given by
\begin{equation}
	\label{momenta1}	
		\pi^a_{\mu}= \frac{\partial \mathcal{L}}{\partial\left(\partial_{0} A_{\mu}^a\right)}=F^a_{0\mu},
 \end{equation}
from which we define the following fundamental Poisson brackets:
\begin{align}
    \left\{A_{\mu}^a(x),A_{\nu}^b(y)\right\}&=\left\{\pi^a_{\mu}(x),\pi^b_{\nu}(y)\right\}=0\\
    \left\{A_{\mu}^a(x),\pi^b_{\nu}(y)\right\}&=\delta^{ab}\delta_{\mu\nu}\delta^{(3)}(x-y)
\end{align}
From equation \eqref{momenta1} we get primary constraint of the model:
\begin{equation}
	\label{priconstraint}	
		\pi^{a}_0=0\equiv T_1^a\approx0,
 \end{equation} 
 The canonical momenta $\pi^a_i$ are given by
 \begin{equation}
	\label{momenta}	
		\pi^a_i=F^a_{0i}.
 \end{equation} 
For the complete investigation of the canonical structure of the model, we must now write down the canonical Hamiltonian of the model, $H_c$, which is obtained from a Legendre transformation that leads to
\begin{equation}
	\label{canHam}	
		H_c = \int d^3x\left\{\frac{1}{2}\pi_i^a\pi_i^a+\pi_i^a\left(\partial_iA_0^a-gf^{abc}A_0^bA_i^c\right)-\frac{1}{4}F_{ij}^aF_{ij}^a-\frac{1}{2}m^2\left(A^a_0A^a_0+A^a_iA^a_i\right)\right\},
 \end{equation}
 From the stability condition of the constraint $T_1^{a}$ \cite {dirac1933, Dirac:1950pj, Dirac:1951zz, Dirac:1958sq, Henneaux:1992ig}, we obtain a secondary constraint
 \begin{align}
     \dot{T}_1^a&=\left\{T_1^{a}(x), H_p(y)\right\}\approx 0\nonumber\\
     &=D_i^{ab}\pi_i^{b}+m^2A_0^{a}\approx0\equiv T_2^a
 \end{align}
where $H_p$ is the primary hamiltonian, namely
\begin{equation}
   H_p=H_c+\int d^3x \lambda^a_1(x)T_1^{a}(x),
  \end{equation}
and  
\begin{equation}
 D_\mu^{ab}=\delta^{ab}\partial_\mu-gf^{abc}A_\mu^c 
\end{equation}
is the covariant derivative in the adjoint representation. The time evolution of the constraint $T^a_2$ determines the Lagrange multiplier $\lambda^a_1$:
\begin{equation}
\lambda^a_1=D^{ab}_iA^b_i.
\end{equation}
This indicates that no more constraints are generated via this iterative procedure. That is, $T_1^a$ and $T_2^a$ are the only constraints of the theory. Further, they form the following second class algebra:
\begin{align}
    \left\{T^a_1(x),T^b_1(y)\right\}&=0,\\
    \left\{T^a_1(x),T^b_2(y)\right\}&=-m^2\delta^{ab}\delta^{(3)}(x-y),\\
    \left\{T^a_2(x),T^b_2(y)\right\}&=gf^{abc}T^c_1\delta^{(3)}(x-y).
\end{align}
The total hamiltonian is then given by
\begin{equation}
\label{totHam}
    H_T=H_c+\int d^3x\left(\lambda_1^aT^a_1+\lambda_2^aT^a_2\right).
\end{equation}
Demanding again time-independence of the constraint $T_1^a$, but this time using the total hamiltonian \eqref{totHam}, the Lagrange multiplier $\lambda_2^a$ is fixed as
\begin{equation}
    \lambda_2^a=\frac{1}{m^2}T_2^a\approx0.
\end{equation}
We are now able to apply the improved gauge unfixing formalism to this model. We have two available choices, as the system presents two second class constraints. By choosing $T_1^a$ as the generator of infinitesimal transformations, the field variations are similar to those of the abelian case \cite{Vytheeswaran:1997jr}. In contrast, the infinitesimal transformations generated by the constraint $T_2^a$ lead to new features in the non-Abelian case. Therefore, we will focus on the second case. We first redefine $T_2^a$ as
\begin{equation}
\chi^a\equiv\frac{1}{m^2}T_2^a,
\end{equation}
such that
\begin{equation}
\left\{\chi^a(x),T_1^b(y)\right\}=\delta^{ab}\delta^{(3)}(x-y).
\end{equation}
$T_1^a$ will be ignored as a constraint. The infinitesimal transformations generated by $\chi^a$ are
\begin{align}
    \delta^aA_0^b(x)&=\int d^3y \varepsilon(y)\{A_0^b(x),\chi^a(y)\}=0,\\
\delta^aA_i^b(x)&=\int d^3y \varepsilon(y)\{A_i^b(x),\chi^a(y)\}=-\frac{1}{m^2}D_i^{ba}\varepsilon(x),\\
\delta^a\pi_0^b(x)&=\delta^a T_1^b(x)=\int d^3y \varepsilon(y)\{\pi_0^b(x),\chi^a(y)\}=-\varepsilon(x)\delta^{ba},\\
\delta^a\pi_i^b(x)&=\int d^3y \varepsilon(y)\{\pi_i^b(x),\chi^a(y)\}=-gf^{abc}\varepsilon(x)\pi_i^c(x)
\end{align}
From the above relations, one can note that $A_i^a$ and $\pi_i^a(x)$ are the relevant noninvariant fields under infinitesimal transformations generated by $\chi^a$. We will start by seeking an invariant field $\tilde{A}_i^a$, which is constructed by a power series of the disregarded constraint $T_1^a$ (\textit{vide} eq. \eqref{F-tilde_1}):
\begin{equation}
\label{A_tilde}
\tilde{A}^a_i(x)=A^a_i(x)+\sum_{k=1}^{\infty}\int\left(\prod_{r=1}^k d^3x_r\right)
c_{ki}^{aw_1...w_k}(x_1,...,x_k;x)\left(\prod_{r=1}^kT^{w_r}_1(x_r)\right).
\end{equation}
The next step is to determine the coefficients $c_{ki}$. For this goal, we impose to $\tilde{A}_i^a$ the variational condition \eqref{F-var}: 
\begin{align}
\delta^m\tilde{A}^a_i(x)=\delta^mA^a_i(x)+&\sum_{k=1}^{\infty}\int\left(\prod_{r=1}^k d^3x_r\right)
\delta^mc_{ki}^{aw_1...w_k}(x_1,...,x_k;x)\left(\prod_{r=1}^kT^{w_r}_1(x_r)\right)+\nonumber\\
+&\sum_{k=1}^{\infty}\int\left(\prod_{r=1}^k d^3x_r\right)c_{ki}^{aw_1...w_k}(x_1,...,x_k;x)\delta^m\left(\prod_{r=1}^kT^{w_r}_1(x_r)\right)=0.
\label{varA}
\end{align}
Equation \eqref{varA} generates an equation for zeroth-order terms in $T_1^a$, from where we find that
\begin{align}
c_{1,i}^{ab}(x,y)&=\left\{A_i^a(x),\chi^b(y)\right\}\nonumber\\
&=\frac{1}{m^2}D_{i,y}^{ba}\delta^{(3)}(x-y),
\label{c1}
\end{align}
where the subscript ‘‘$y$’’ in the covariant derivative indicates in which space-time variable it is acting on. From the linear equation in $T_1^{a}$, we can find the second coefficient of \eqref{A_tilde}:
\begin{align}
c_{2,i}^{abc}(x,y,z)&=\frac{1}{2}\left\{c_{1,i}^{ac}(x,y),\chi^b(z)\right\}\nonumber\\
&=\frac{gf^{cad}}{2m^4}\delta^{(3)}(x-y)D^{bd}_{iz}\delta^{(3)}(y-z).
\label{c2}
\end{align}
Performing the same procedure for second-order terms, we arrive at the third coefficient of the series \eqref{A_tilde}
\begin{align}
c_{3,i}^{abcd}(x,y,z,w)&=\frac{1}{3}\left\{c_{2,i}^{acd}(x,y,z),\chi^b(w)\right\}\nonumber\\
&=\frac{g^2f^{dae}f^{ceg}}{3!m^6}\delta^{(3)}(x-y)\delta^{(3)}(y-z)D_{i,w}^{bg}\delta^{(3)}(z-w).
\label{c3}
\end{align}
From equations \eqref{c1}, \eqref{c2} and \eqref{c3}, one concludes that, for $k\geq4$, the coefficients $c_k$ have the following form:
\begin{align}
c_{k,i}^{aw_1...w_k}=&\frac{g^{k-1}}{k!m^{2k}}f^{w_kab_1}f^{w_{k-1}b_1b_2}\ldots
f^{b_{k-2}w_3b_{k-1}}f^{b_{k-1}w_2v_k}\times
\nonumber\\
&\times \delta^{(3)}\left(x-x_1\right)\left(\prod_{j=2}^{k-1}\delta^{(3)}\left(x_j-x_{j+1}\right)\right)D_{i,x_k}^{w_1v_k}\delta^{(3)}(x_{k-1}-x_k)
\end{align}
Hence, the general form of the field $\tilde{A}^a_i$ is given as follows:
\begin{align}
\tilde{A}^a_i(x)&=A^a_i(x)-\frac{1}{m^2}D_{i}^{ba}T_1^b(x)
+\sum_{k=2}^{\infty}\int\left(\prod_{r=1}^k d^3x_r\right)
c_i^{a,w_1...w_k}(x_1,...,x_k;x)\left(\prod_{r=1}^kT^{w_r}_1(w_r)\right)\nonumber\\
&=A^a_i(x)-\frac{1}{m^2}D_{i}^{ba}\pi_0^b(x)
-\sum_{k=2}^{\infty}\frac{g^{k-1}}{k!m^{2k}}F^{w_ka...w_2c}
 \left(\prod_{r=2}^{k}\pi^{w_r}_0(x)\right)
\left(D_i^{cw_1}\pi_0^{w_1}(x)\right)
\end{align}
where
\begin{equation}
F^{w_ka...w_2c}=f^{w_kab_1}f^{w_{k-1}b_1b_2}\ldots
f^{b_{k-2}w_3b_{k-1}}f^{b_{k-1}w_2c}.
\end{equation}
Writing the series for $\tilde{A}_i^{a}$ explicitly, we have
\begin{eqnarray}
\label{SS}
    \tilde{A}^{a}_i=A^a_i-\frac{1}{m^2}D_{i}^{ba}\pi_{0}^b-\frac{gf^{cad}}{2m^4}\pi^{b}_{0}D^{bd}_i\pi^{c}_{0}-\frac{g^{2}f^{dae}f^{ceg}}{3!m^{6}}\pi_0^{b}\pi_0^{c}D_i^{bg}\pi_0^{d}\ldots,
\end{eqnarray}
which can also be written as
\begin{equation}
\tilde{A}^a_i = A^a_i -D_{i}^{ba}\xi^b-\frac{g}{2}f^{cad}\xi^bD_i^{bd}\xi^c-\frac{g^2}{6}f^{dac}f^{ceg}\xi^{b}\xi^{c}D_i^{bg}\xi^{d}\ldots,
\label{stueckelbergA}
\end{equation}
where $\xi^{a}\equiv \frac{\pi^{a}_0}{m^{2}}$. If we identify $\xi^a$ as a Stueckelberg field \cite{Stueckelberg:1957zz, Ruegg:2003ps}, then the expression \eqref{stueckelbergA} is the same as the obtained in \cite{Capri:2016ovw, Capri:2016jgn}, in which they have considered a field $h=e^{ig\xi^aS^a}$ (with $S^a$ being the generators of the gauge group SU(N)) acting on $A_{\mu}^a$, in order to obtain an action put in a local form.

The fields $\tilde{\pi}_i^a$ can be obtained using the same algorithm applied for the fields $\tilde{A}^{a}_i$. Then, we present the first three coefficients of the series \eqref{F-tilde_1} for the fields $\tilde{\pi}_i^a$, which are:
\begin{align}
\label{3firstB}
b_{1,i}^{ab}(x,y)&=\frac{1}{m^2}g f^{bca}\pi^c_i(y)\delta^3(x-y),\nonumber\\
b_{2,i}^{abc}(x,y,z)&=\frac{1}{2m^4}g^2f^{cda}f^{bed}\pi^e_i(z)\delta^3(x-y)\delta^3(y-z),\\
b_{3,i}^{abcd}(x,y,z,w)&=\frac{1}{6m^6}g^3f^{dea}f^{cge}f^{bhg}\pi^h_i(w)\delta^3(x-y)\delta^3(y-z)\delta^3(z-w),\nonumber
\end{align}
by which, with a little observation, it is possible to build up a general form of the coefficients for all $k\geq4$:
\begin{align}
\label{nfirstB}
b_{k,i}^{a,w_1,...,w_k}(x,y,z,w)=&\frac{g^k}{k!m^{2k}} f^{w_kb_{k-1}a}f^{w_{k-1}b_k b_{k-1}}\ldots f^{w_1 v_k b_k}\times\nonumber\\
&\times\pi^{v_k}_i(x_k)\delta^3(x-x_1)\left(\prod_{j=2}^{k-1}\delta^3(x_j-x_{j+1}) \right).
\end{align}
Hence, the results of \eqref{3firstB} and \eqref{nfirstB} allow us to obtain the expression for the GU-corrected variable $\tilde{\pi}_i^{a}$:
\begin{align}
    \label{correctedpi}
    \tilde{\pi}_i^{a}=\pi_i^{h}\left[\delta^{ha}+gf^{bha}\xi^b+\frac{g^2}{2!}f^{cda}f^{bhd}\xi^b \xi^c+\frac{g^3}{3!}f^{dea}f^{cge}f^{bhg}\xi^b\xi^c\xi^d +\ldots\right],
\end{align}
where $\xi^a\equiv \frac{\pi^{a}_0}{m^2}$. One can now check that the new fields satisfy $\left\{\tilde{A}_{i}^a,\chi^b \right\}=0$ and $\left\{\tilde{\pi}_{i}^a,\chi^b \right\}=0$, \textit{i.e.}, $\tilde{A}_i^a$ and $\tilde{\pi}^{a}_i$ are first class variables. By replacing \eqref{stueckelbergA} and \eqref{correctedpi} into \eqref{canHam} and \eqref{act1}, one can obtain the first class Hamiltonian and the first class Lagrangian, respectively.

\section{The $SU(2)$ Skyrme Model}

We now consider the $SU(2)$ Skyrme model \cite{Skyrme:1961vq,Adkins:1983ya,Oliveira:1998ek}, that describes baryons and their interactions through soliton solution from the Lagrangian
\begin{equation}
\label{skirmeL1}
    L=\int d^3r\left[-\frac{F_{\pi}^2}{12}\Tr{\partial_iU\partial_iU^+}+\frac{1}{32e^2}\Tr{\left[U^+\partial_iU,U^+\partial_jU\right]}\right],
\end{equation}
where $F_{\pi}$ is the pion decay constant, $e$ is a dimensionless parameter and $U$ is an $SU(2)$ matrix. By inserting a collective coordinate $A(t)$ into the Lagrangian above and substituting $U(r)$ by $U(r,t)=A(t)U(r)A^+(t)$, we get\footnote{Technical details can be found in \cite{Adkins:1983ya}. The crucial point to be addressed here is the $SU(2)$ group has a  Wess-Zumino term that vanishes, so it is triangle anomaly-free and the Hamiltonian admits a diagonal form.}
\begin{equation}
L=-M+\lambda\Tr{\partial_0A\partial_0A^{-1}},
\end{equation}
where $M$ is the soliton mass, which, in the Skyrme ansatz $U(r)=exp\left(iF(r)\bm{\tau}\cdot\bm{\hat{r}}\right)$, is given by
\begin{equation}
    M=4\pi\int_0^{\infty}drr^2\left\{\frac{1}{8}F_{\pi}^2\left[\left(\frac{\partial F}{\partial r}\right)^2+2\frac{\sin^2F}{r^2}\right]+\frac{1}{2e^2}\frac{\sin^2F}{r^2}\left[\frac{\sin^2 F}{r^2}+2\left(\frac{\partial F}{\partial r}\right)^2\right]\right\};
    \end{equation}
$\lambda=2/3\left(1/e^3F_{\pi}\right)\Lambda$, with
\begin{equation}
\Lambda=\int_0^{\infty}dxx^2\sin^2F\left\{1+4\left[\left(\frac{\partial F}{\partial x}\right)^2+\frac{\sin^2F}{x^2}\right]\right\},
\end{equation}
in which $x$ is a dimensionless variable defined by $x=eF_{\pi}r$.

The $SU(2)$ matrix $A$ can be written as $A=a_0+i\bm{a}\cdot\bm{\tau}$ with the constraint
\begin{equation}
\label{primary}    
    T_1=a_ia_i-1\approx 0, \quad i=0,1,2,3.
\end{equation}
Then, the Lagrangian (\ref{skirmeL1}) can be rewritten as a function of the $a_{i}$'s as
\begin{equation}
\label{skirmeL2}
L=-M+2 \lambda \dot{a}_{i} \dot{a}_{i}
\end{equation}
In order to pass to the Hamiltonian formalism, we compute the canonical momenta, which are given by
\begin{equation}
\label{mometask}
\pi_{i}\equiv\frac{\partial L}{\partial \dot{a}_{i}}=4\lambda \dot{a}_{i},
\end{equation}
Through the Legendre transformation we get the canonical Hamiltonian

\begin{equation}
\label{HcanS}
H_c=M+\frac{1}{8\lambda}\pi_{i}\pi_{i}.
\end{equation}
From the consistency condition that the constraint $T_1$ cannot evolve in time, we obtain a secondary constraint
\begin{equation}
\label{secondary}
T_2=a_i \pi_i \approx 0.
\end{equation}
Then, the total Hamiltonian is
\begin{equation}
    H_T=M+\frac{1}{8\lambda}\pi_{i}\pi_{i}+u_1\left(a_ka_k-1\right)+u_2a_i\pi_i.
    \label{TotHamSkyrme1}
\end{equation}
The conservation of the constraint $T_2$ determines the Lagrange multiplier $u_1$:
\begin{equation}
   u_1=\frac{\pi_i\pi_i}{8\lambda a_ja_j},
\end{equation}
that confirms the system possesses only two constraints, and they are second class, with Poisson algebra given by
\begin{equation}
\label{paircan}
\left\{T_a,T_b\right\}=2\epsilon_{ab}a_ia_i, \quad a,b=1,2,
\end{equation}
where $\epsilon_{ab}=1$. The total Hamiltonian can now be rewritten as
\begin{equation}
    H_T=M+\frac{1}{8\lambda}\pi_{i}\pi_{i}+\frac{\pi_i\pi_i}{8\lambda a_ja_j}\left(a_ka_k-1\right)+u_2a_i\pi_i.
    \label{TotHamSkyrme2}
\end{equation}
The time evolution of the constraint $T_1$ fixes the Lagrange multiplier $u_2$:
\begin{equation}
    u_2=-\frac{a_i\pi_i}{4\lambda a_ja_j},
\end{equation}
and the total Hamiltonian \eqref{TotHamSkyrme2} becomes
\begin{equation}
    H_T=M+\frac{1}{8\lambda}\left(2-\frac{1}{a_ja_j}\right)-\frac{\left(a_i\pi_i\right)^2}{4\lambda_ja_ja_j}.
    \label{TotHamSkyrme3}
\end{equation}

The second class nature shown by the constraints in equation \eqref{paircan} allows us applying the improved gauge unfixing formalism. Now we need to choose one of the constraints to be the generator of infinitesimal transformations. The two possibilities can be analysed separately.

\subsubsection*{$T_1$ as first class constraint}

In this case, we choose the constraint (\ref{primary}) as the generator of infinitesimal transformations, $\chi$. We can redefine it so that the two constraints form an approximate canonically conjugate pair. Thus, we reclassify $T_1$ as
\begin{equation}
\chi=\frac{T_{1}}{2 a_{i} a_{i}}=\frac{a_{i} a_{i}-1}{2 a_{j} a_{j}} .
\end{equation}
The constraint $T_2$ will be discarded. Therefore, the infinitesimal variations of the relevant quantities are given by
\begin{equation}
\label{varpi}
\begin{aligned}
\delta a_{i}&=\varepsilon\left\{a_{i}, \chi\right\}=0. \\
\delta \pi_{i}&=\varepsilon\left\{\pi_{i}, \chi\right\}=\varepsilon\left(-\frac{a_i}{a_ja_j}+\frac{2a_i\chi}{a_ja_j}\right) \\
\delta T_2&=\varepsilon\left\{T_2, \chi\right\}=-\varepsilon \left(1+2\chi\right)
\end{aligned}
\end{equation}
As presented in sec. \ref{GeneralGU}, the first class variables are constructed as a power series of the disregarded constraint, namely
\begin{equation}
\begin{aligned}
&\tilde{a}_i=a_i+\sum_{n=1} b_nT_2^n,\\
&\tilde{\pi}_i=\pi_i+\sum_{n=1} c_nT_2^n,
\end{aligned}
\end{equation}
where the coefficients $b_n$ and $c_n$ are calculated from (\ref{con-n}). By using the expression (\ref{varpi}), we can see that, for the variable $\tilde{a}_i$, all coefficients $b_n$ are trivially zero. For $\tilde{\pi}_i$, we obtain
\begin{equation}
c_1=\frac{\left(2\chi-1\right)a_i}{2a_ja_j-1}.
\end{equation}
Because $\delta c_1=0$, all coefficients are null for $n\geq2$. Therefore, the first class variables are
\begin{equation}
\begin{aligned}
    \tilde{a}_i&=a_i\\
    \tilde{\pi}_i&=\pi_i+\frac{\left(2\chi-1\right)a_i}{2a_ja_j-1}a_k\pi_k.
    \label{fcpi}
\end{aligned}
\end{equation}
The construction of the Hamiltonian through the new variables is essential to verify if the new system is now of the first class type. We can verify this through the Poisson brackets between the symmetry generator $\chi$ and the new Hamiltonian. By the insertion of (\ref{fcpi}) in the Hamiltonian (\ref{HcanS}), we obtain the following first-class Hamiltonian:
\begin{align}
    \tilde{H_c}&=M+\frac{1}{8\lambda}\tilde{\pi}_i\tilde{\pi}_i\nonumber\\
             &=H_c+\frac{1}{4\lambda}\frac{\left(2\chi-1\right)\left(a_i\pi_i\right)^2}{2a_ja_j-1}+\frac{1}{8\lambda}\left[\frac{\left(2\chi-1\right)a_ia_j\pi_j}{2a_ka_k-1}\right]^2.
             \label{Hinv1}
\end{align}
It can now be verified that $\left\{\chi,\tilde{H}_c\right\}=0$, which means that $\chi$ and $\tilde{H_c}$ describe a gauge theory in this new system \cite{Henneaux:1992ig}.

\subsubsection*{$T_2$ as first class constraint}

We will now consider the other choice of generator of infinitesimal transformations, \textit{i.e.}, we will now rescale $T_2$ as
\begin{equation}
    \chi^{\prime}=-\frac{T_{2}}{2a_ia_i}=-\frac{a_i\pi_i}{2a_ja_j},
\end{equation}
and $T_1$ will be ignored as a constraint. $\chi^{\prime}$ generates infinitesimal transformations as follows
\begin{equation}
\label{varpi2}
\begin{aligned}
\delta a_{i}&=-\varepsilon\left\{a_{i}, \chi^{\prime}\right\}=- \frac{\varepsilon a_i}{2a_ja_j}\\
\delta \pi_{i}&=\varepsilon\left\{\pi_{i}, \chi^{\prime}\right\}=\frac{\varepsilon \pi_i}{2a_ja_j}+\frac{2\varepsilon\chi^{\prime} a_i}{a_k a_k}\\
\delta T_1&=\varepsilon\left\{T_1, \chi^{\prime}\right\}=-\varepsilon.
\end{aligned}
\end{equation}
So, both $a_i$ and $\pi_i$  are second class variables and must be redefined. The first class variables $\tilde{a}_i$ and $\tilde{\pi}_i$ are
\begin{equation}
\label{seriesApi}
\begin{aligned}
&\tilde{a}_i=a_i+\sum_{n=1} b_nT_1^n,\\
&\tilde{\pi}_i=\pi_i+\sum_{n=1} c_nT_1^n.
\end{aligned}
\end{equation}
For both variables, we have an infinite number of coefficients, because the series do not truncate. For $\tilde{a}_i$, the first two coefficients are
\begin{equation}
\begin{aligned}
&b_1=- \frac{a_i}{2a_ja_j},\\
&b_2=-\frac{a_i}{8\left(a_ja_j\right)^2}.
\end{aligned}
\end{equation}
For $n\geq3$, the general relation is
\begin{equation}
    b_n=-\frac{\frac{1}{2}\left(\frac{1}{2}-1\right)\ldots\left(\frac{1}{2}-n+1\right)}{n!}\frac{a_i}{\left(a_ja_j\right)^n}.
\end{equation}
Therefore, $\tilde{a}_i$ becomes
\begin{equation}
    \tilde{a}_i=a_i\left[1-\frac{(a_ja_j-1)}{2a_ka_k}-\frac{(a_ja_j-1)^2}{8\left(a_ka_k\right)^2}+\ldots\right],
\end{equation}
which can be put in a closed form as
\begin{equation}
\tilde{a}_i=a_i\left[1-\frac{\left(a_ja_j-1\right)}{a_ka_k}\right]^{\frac{1}{2}}=\frac{a_i}{(a_j a_j)^\frac{1}{2}}.   
\end{equation}

Now, for the variable $\tilde{\pi}_i$, the first two coefficients are
\begin{equation}
\begin{aligned}
&c_1=\frac{\pi_i}{2a_ja_j}+\frac{2a_i\chi^{\prime}}{a_ka_k},\\
&c_2=\frac{3\pi_i}{8\left(a_ja_j\right)^2}+\frac{a_i\chi^{\prime}}{2\left(a_k a_k\right)^2},
\end{aligned}
\end{equation}
By replacing the $c_n$ in the expression (\ref{seriesApi}), the variable $\tilde{\pi}_i$ becomes
\begin{equation}
    \label{tildepic2}
    \tilde{\pi}_i=\pi_i+\left(\frac{\pi_i}{2a_ka_k}+\frac{2a_i\chi^{\prime}}{a_ka_k}\right)(a_j a_j-1)+\left(\frac{3\pi_i}{8\left(a_ka_k\right)^2}+\frac{a_i\chi^{\prime}}{2(a_k a_k)^2}\right)(a_j a_j-1)^2\ldots \;.
\end{equation}
Unlike $\tilde{a}_i$, one cannot obtain here a closed form to the first class variable $\tilde{\pi}_i$. We would like to emphasize that this is a different result from the one obtained in \cite{vytheeswaran2002hidden}. 

By the insertion of (\ref{tildepic2}) in the Hamiltonian (\ref{HcanS}), we obtain a new first class Hamiltonian:
\begin{equation}
    \label{Hinv2}
    \tilde{H}^{\prime}_c=M+\frac{1}{8\lambda}\tilde{\pi}_i\tilde{\pi}_i
\end{equation}
It is possible to verify that $\left\{\chi, \tilde{H}^{\prime}_c\right\}=0$.
\section{Conclusions}
\label{Conclusions}

In this paper, we have extended the results of the improved GU formalism \cite{Neto:2006gt,Neto:2009rm} to take into account the presence of non-Abelian variables, i.e., variables with values in the Lie algebra of a non-commutative group. The central point of this formalism is the gauge invariance restitution of dynamical systems with second-class constraints. Following an iterative procedure given in the previous sections, the second-class variables can be converted to gauge invariant quantities and then we have a gauge invariant system. In case a system has two second-class constraints, there are two ways to define the GU variables, depending on the constraint we have initially chosen to be the gauge symmetry generator.
In particular, we have investigated the gauge invariance restitution for the massive $SU(N)$ Yang-Mills model and for the $SU(2)$ Skyrme model. In the former, the proper choice of the gauge symmetry generator allows us to reproduce exactly the infinitesimal form of the gauge variations for the case of massless Yang-Mills. The obtained expansions for the GU variables, written in terms of a  power series of the discarded constraint,  reproduce exactly the (non-polynomial) Stückelberg construction derived in previous works. 
On the other hand, we have obtained two different gauge theories for the SU(2) Skyrme model. For the first case ($T_1$ as a generator, Eq. (\ref{primary})), the GU variables acquire a simpler form comparing with the second case ($T_2$ as a generator, Eq. (\ref{secondary})), where the respective GU variables have an infinite number of coefficients. However, these GU variables could be written in a closed form. Both systems go back to the original second-class one when we impose the initial boundary conditions. Finally, we can mention that for the massive Yang-Mills model and for the SU(2) Skyrme model, the Poisson brackets between the GU variables are equals to the Dirac brackets between the original second-class variables \cite{Neto:2009rm}. These results can indicate the validate of the consistency of our formalism.


\section*{Acknowledgements}

The Coordenação de Aperfeiçoamento de Pessoal de Nível Superior (CAPES) and the Fundação de Amparo à Pesquisa do Estado de Minas Gerais (FAPEMIG) are ackknowledged for financial support.
E. M. C. Abreu e Jorge Ananias Neto thank CNPq (Conselho Nacional de Desenvolvimento Cient\' ifico e Tecnol\'ogico), Brazilian scientific support federal agency, for partial financial support, CNPq-PQ, Grants numbers  406894/2018-3 (Everton M.C. Abreu), and 307153/2020-7 (Jorge Ananias Neto). 


\bibliography{biblio}
\end{document}